\documentclass[page-classic]{epl2} 

\usepackage{pxfonts}

\title{Relativistic Equilibrium Distribution by Relative Entropy Maximization}
\shorttitle{Relativistic Equilibrium Distribution}

\author{Tadas K. Nakamura} 

\institute{                    
   Fukui Prefectural University, 910-1195 Fukui, JAPAN
}
\pacs{02.50.Cw}{Probability theory}
\pacs{05.20.Dd}{Kinetic theory}
\pacs{51.10.+y}{Kinetic and transport theory of gases}

\abstract{
The equilibrium state of a relativistic gas has been calculated based
on the maximum entropy principle. Though the relativistic equilibrium
state was long believed to be the J\"uttner distribution, a number
of papers have been published in recent years proposing alternative
equilibrium states. However, some of these papers do not pay enough
attention to the covariance of distribution functions, resulting confusion
in equilibrium states. Starting from a fully covariant expression
to avoid this confusion, it has been shown in the present paper that
the J\"uttner distribution is the maximum entropy state if we assume
the Lorentz symmetry.
}

\begin{document}

\maketitle

\section{Introduction}

Little after the establishment of the theory of relativity, the equilibrium
particle distribution of a relativistic gas was investigated. The
distribution obtained, which is called J\"uttner distribution \cite{juttner11,synge57},
has been long and widely believed. However, relatively recent years a number
of papers have been published proposing equilibrium distribution functions
other than the J\"uttner distribution (\cite{horwitz81,horwitz89,lehmann06,schieve05,dunkel07}
and references therein). Dunkel and coworkers \cite{dunkel07,cubero09}
have examined the discrepancy in the equilibrium distributions as
the maximum entropy state, and showed that the difference comes from
the choice of the reference measure.

The maximum entropy state cannot be uniquely determined when one naively
defines the entropy such as $S=-\int f(\mathbf{x},\mathbf{v})\ln f(\mathbf{x},\mathbf{v})\, d\mathbf{x}d\mathbf{v}$
(symbols have conventional meaning in the present paper unless otherwise
stated). For instance, the result would be different if we rewrite
distribution function as a function of momentum $\mathbf{p}$ instead
of velocity $\mathbf{v}$. To overcome this difficulty, it was proposed
in Ref \cite{dunkel07} to maximize the following relative entropy 
\begin{equation}
S=-\int f(\mathbf{x},\mathbf{v})\ln f(\mathbf{x},\mathbf{v})/\rho(\mathbf{x},\mathbf{v})\, d\mathbf{x}d\mathbf{v}\,,
\label{eq:relent}
\end{equation}
based on a given reference measure $\rho$.  In the above expression
$f$ is the phase space distribution of particles and $\rho$ is the
reference measure \cite{dunkel07}.  In this paper we denote a three
vector by a bold font (e.g., $\mathbf{x}$) and a four vector by an
upper bar (e.g., $\bar{x}$). Each component of a vector is represented
by a subscript or a superscript (e.g., $x_{\mu}$ or $x^{\mu}$).

The equilibrium distribution is uniquely determined by maximizing
the relative entropy once the reference measure is given. The mathematical
procedure in this approach is essentially the same as the one utilized
in Ref \cite{synge57} to derive the J\"uttner distribution. What
is called {}``a priori probability'' in Ref \cite{synge57} plays
the same role as the reference measure in Ref \cite{dunkel07}.

Two possibilities for the reference frame were suggested in Ref
\cite{dunkel07}. One is the constant distributions a function 
of momentum, 
and the J\"ttner distribution is obtained from this measure.
This calculation is essentially the same as the one in 
Ref\cite{synge57}. 
Another possibility 
suggested in Ref \cite{dunkel07} which is inversely proportional 
to the energy.
It was argued this measure is derived from the Lorentz
symmetry in Ref \cite{dunkel07} and the result is the alternative
equilibrium distribution proposed in recent papers. However, as we
will see in the present paper, there is a confusion on the
relativistic phase space density in this argument. The Lorentz
invariant reference measure is the same as the one in Ref
\cite{synge57}, i.e., the constant measure, which gives the J\"uttner
distribution.

\bigskip{}

There is a misleading point in defining a phase space density such
as a particle distribution in relativity. When we express a phase
space density as the time evolution of the density in a six (three
space + three momentum) dimensional phase space, it appears to be
a Lorentz invariant. Actually, it can be proved \cite{vankampen69}
(see also \cite{synge57,debbasch01}) that $f(t,\mathbf{x},\mathbf{p})=f(t',\mathbf{x'},\mathbf{p}')$
when the two sets of coordinates $(t,\mathbf{x},\mathbf{p})$ and
$(t',\mathbf{x}',\mathbf{p}')$ are related by the Lorentz transform,
in other words, they are the same point in the spacetime denoted by
different reference coordinates. However, this does not mean $f(t,\mathbf{x},\mathbf{p})d\mathbf{x}d\mathbf{p}=f(t',\mathbf{x'},\mathbf{p}')d\mathbf{x}'d\mathbf{p}'$
because $\mathbf{x}$ and $\mathbf{x}'$ do not belong to the same
spatial volume. In this sense, phase a space density in the form of
$f(t,\mathbf{x},\mathbf{p})$ is not covariant but frame dependent.
It seems that some of recent papers do not pay enough attention to
this fact, resulting confision in treating Lorentz transfrom.

In the present paper, we examine this confusing point by starting
from the fully covariant distribution function proposed by Hakim \cite{hakim67},
and the result shows the reference measure should be constant to satisfy
the full Lorentz symmetry; the one introduced in Ref \cite{dunkel07}
is invariant under the Lorentz transform only in the momentum space.
This result means the maximum entropy state with Lorentz symmetry
must be the J\"uttner distribution.

\section{Relativistic Phase Space Density}

Let us suppose a relativistic gas as an example. The conservation
law of its particle number is expressed in the form of flux divergence
in relativity:\begin{equation}
\frac{\partial}{\partial x_{\mu}}\, J_{\mu}=0\,,\end{equation}
where \begin{equation}
J_{\mu}=n_{0}u_{\mu}\label{eq:flux}\,.\end{equation}
 is the four flux derived from the proper number density $n_{0}$
and the four velocity of the matter $u_{\mu}$. When we split the
spacetime as $t_{\Sigma}=x_{\Sigma0}$ and $\mathbf{x_{\Sigma}}=(x_{\Sigma1,}x_{\Sigma2},x_{\Sigma3})$
by choosing a specific reference frame $\Sigma$, the above conservation
is written as
\begin{equation}
\frac{\partial}{\partial t}n_{\Sigma}(t_{\Sigma},\mathbf{x}_{\Sigma})+\nabla\mathbf{J}_{\Sigma}(t_{\Sigma},\mathbf{x}_{\Sigma})=0\,.\label{eq:flux2}\end{equation}
In the above expression, $n_{\Sigma}=J_{\Sigma0}$ and $\mathbf{J}_{\Sigma}=(J_{\Sigma1},J_{\Sigma2},J_{\Sigma3})$
are the number density and flux in the three dimensional space; the
subscript $\Sigma$ is to explicitly express the frame dependence.

When we decompose the spacetime in another reference frame $\Sigma'$,
obviously $n_{\Sigma'}$ is different from $n_{\Sigma}$. Moreover,
$n_{\Sigma}$ and $n_{\Sigma'}$ cannot be related with a Jacobian
$\partial\mathbf{x}_{\Sigma}/\partial\mathbf{x}_{\Sigma'}$ as\begin{equation}
n_{\Sigma}d\mathbf{x}_{\Sigma}=n_{\Sigma'}\frac{\partial\mathbf{x}_{\Sigma}}{\partial\mathbf{x}_{\Sigma'}}\, d\mathbf{x}_{\Sigma'}\,,\end{equation}
because $\mathbf{x}_{\Sigma}$ and $\mathbf{x}_{\Sigma}$ belong to
different spacelike volumes. There is no function to relate $\mathbf{x}_{\Sigma}$
and $\mathbf{x}_{\Sigma'}$ as
$\mathbf{x}_{\Sigma}=\mathbf{X}(\mathbf{x}_{\Sigma'})$ where $\mathbf X$ that does not depend on the time coordinate.
\bigskip{}

The above argument on the number density in a three dimensional space
is also valid for phase space densities in a six dimensional space.
A phase space density is often expressed as $f(t,\mathbf{x},\mathbf{p})$
and it should be denoted in our notation as $f_{\Sigma}(t_{\Sigma},\mathbf{x}_{\Sigma},\mathbf{p}_{\Sigma})$
because the expression is based on a specific choice of the reference
frame like $n_{\Sigma}$ in (\ref{eq:flux2}). However, it is
generally believed that the phase space density is unchanged under
the Lorentz transform. This is true in the sense that the value of
the phase space density is unchanged \cite{synge57,vankampen69,debbasch01},
but $f_{\Sigma}(t_{\Sigma},\mathbf{x}_{\Sigma},\mathbf{p}_{\Sigma})$
is defined only on a space volume in a specific reference frame, and
not directly applicable to other reference frames. 

To correctly treat the phase space density, we derive the frame-dependent
phase space density $f_{\Sigma}(t_{\Sigma},\mathbf{x}_{\Sigma},\mathbf{p}_{\Sigma})$
from the fully covariant expression proposed by Hakim \cite{hakim67}.
The relativistic particle distribution $N(\bar{x},\bar{p})$ is defined
such that $\bar{j}$ in the following expression becomes the particle
four-current:\begin{equation}
j_{\mu}(\bar{x})=\int d_{4}p\,2mu_{\mu}N(\bar{x},\bar{p})\,\theta(p^{0})\delta(p^{\mu}p_{\mu}-m^{2})\,,\end{equation}
where $\theta$ and $\delta$ are the theta and delta functions, and
$m$ is the particle rest mass. 

In the above expression, $N(\bar{x},\bar{p})$ can be interpreted
as the proper density of the fluid element that has the four velocity
$\bar{u}=\bar{p}/m$, just like $n_{0}$ in (\ref{eq:flux}). Thus
its covariant form must be a four vector, which is expressed as $N(\bar{x},\bar{p})\bar{u}$,
like $\bar{J}$ in (\ref{eq:flux}). The delta function is due to
the energy shell and the theta function is to discard the negative
energy solution. Hakim \cite{hakim67} has introduced the above expression
for the distribution of particle number, however, it is generally
valid for a conserved density flowing with the four velocity $\bar{u}$,
therefore, it can be applied to a probability distribution or a reference
measure to calculate entropy in the following.

When we pick up one reference frame $\Sigma$ and denote its unit
vectors in each coordinate direction as $(\bar{e}_{\Sigma t},\bar{e}_{\Sigma x},\bar{e}_{\Sigma y},\bar{e}_{\Sigma z})$,
an arbitrary point in the eight dimensional phase space $(\bar{x},\bar{p})$
can be represented in this reference frame as
\begin{equation}
t_{\Sigma}=e_{\Sigma t}^{\mu}x_{\mu},~~\mathbf{x}_{\Sigma}=(e_{\Sigma
  x}^{\mu}x_{\mu},e_{\Sigma y}^{\mu}x_{\mu},e_{\Sigma z}^{\mu}x_{\mu})
\,,\end{equation}
and
\begin{equation}
E_{\Sigma}=e_{\Sigma t}^{\mu}p_{\mu},~~\mathbf{p}_{\Sigma}=(e_{\Sigma x}^{\mu}p_{\mu},e_{\Sigma y}^{\mu}p_{\mu},e_{\Sigma z}^{\mu}p_{\mu})\,.\end{equation}
A frame-dependent phase space density $f_{\Sigma}(t_{\Sigma},\mathbf{x}_{\Sigma},\mathbf{p}_{\Sigma})$
is then calculated from $N(\bar{x},\bar{p})$ as \begin{eqnarray}
f_{\Sigma}(t_{\Sigma},\mathbf{x}_{\Sigma},\mathbf{p}_{\Sigma}) & = & 2m\int e_{\Sigma t}^{\mu}u_{\mu}N(\bar{X},\bar{P})\,\theta(p^{0})\delta(E_{\Sigma}^{2}-\mathbf{p}_{\Sigma}^{2}-m^{2})\, dE_{\Sigma}\nonumber \\
\, & = & \frac{me_{\Sigma t}^{\mu}u_{\mu}}{E_{\Sigma}}\, N(\bar{X},\bar{P})=N(\bar{X},\bar{P})\,,\label{eq:phi}\end{eqnarray}
where $\bar{u}=\bar{p}/m$, and $\bar{X}$ and $\bar{P}$ are the
covariant expression of the four dimensional position and momentum
correspond to $(t,\mathbf{x}_{\Sigma},\mathbf{p}_{\Sigma})$, 
\begin{equation}
\bar{X}=t_{\Sigma}\bar{e}_{\Sigma t}+x_{\Sigma}\bar{e}_{\Sigma x}+y_{\Sigma}\bar{e}_{\Sigma y}+z_{\Sigma}\bar{e}_{\Sigma z}\,,\end{equation}
and
\begin{equation}
\bar{P}=\sqrt{\mathbf{p}^{2}+m^{2}}\,\bar{e}_{\Sigma t}+p_{\Sigma x}\bar{e}_{\Sigma x}+p_{\Sigma y}\bar{e}_{\Sigma y}+p_{\Sigma z}\bar{e}_{\Sigma z}\,.\end{equation}
From (\ref{eq:phi}) van Kampen \cite{vankampen69} concluded that
$f$ is unchanged under the Lorentz transform (his derivation is different
from ours, but the result is the same). He considered the above result
is purely kinematical. It is true in the sense that no equation of
motion is required for (\ref{eq:phi}), however, it implicitly includes
kinetics in the expression of the energy shell. For example, if the
relativistic kinetics were such that the energy shell is expressed
as $4m^{3}\delta(E_{\Sigma}^{4}-\mathbf{p}_{\Sigma}^{4}-m^{4})$,
(\ref{eq:phi}) would be\begin{equation}
f_{\Sigma}(t_{\Sigma},\mathbf{x}_{\Sigma}\mathbf{p}_{\Sigma})=\frac{m^{3}e_{\Sigma t}^{\mu}u_{\mu}}{E_{\Sigma}^{3}}\, N(\bar{X},\bar{P})=\frac{m^{2}}{E_{\Sigma}^{2}}\, N(\bar{X},\bar{P})\,,\end{equation}
which means the value of $f$ changes under the Lorentz transform.
This example demonstrates the fact that $f_{\Sigma}$ is not identical
to $N$, but should be derived from $N$.

\section{Lorentz Invariant Reference Frame}

In (\ref{eq:phi}) we assumed the spatial coordinates $(t_{\Sigma},\mathbf{x}_{\Sigma})$
and the momentum coordinates $(E_{\Sigma},\mathbf{p}_{\Sigma})$ are
defined in the same reference frame $\Sigma$. Mathematically the
reference frames to define spatial and momentum coordinates do not
have to be the same; we may have a phase space density whose spatial
coordinates are defined in $\Sigma$ and momentum coordinates are
in $\Sigma'$ as in the following form:\begin{eqnarray}
f_{\Sigma\Sigma'}(t_{\Sigma},\mathbf{x}_{\Sigma},\mathbf{p}_{\Sigma'}) & = & 2m\int e_{\Sigma t}^{\mu}u_{\mu}N(\bar{X},\bar{P}')\,\theta(p^{0})\delta(E_{\Sigma'}^{2}-\mathbf{p}_{\Sigma'}^{2}-m^{2})\, dE_{\Sigma'}\nonumber \\
\, & = & \frac{me_{\Sigma t}^{\mu}u_{\mu}}{E_{\Sigma'}}\, N(\bar{X},\bar{P}')\,,\label{eq:phi2}\end{eqnarray}
with
\begin{equation}
\bar{P}'=E_{\Sigma'}\bar{e}_{\Sigma't}+p_{\Sigma'x}\bar{e}_{\Sigma'x}+p_{\Sigma'y}\bar{e}_{\Sigma'y}+p_{\Sigma'z}\bar{e}_{\Sigma'z}\,.
\end{equation}
From the above expression we understand that the factor of $e_{\Sigma t}^{\mu}u_{\mu}$
comes from the spatial Lorentz transform whereas the factor of $1/E_{\Sigma'}$
is due to the transform in the momentum space. They are canceled out
when $\Sigma=\Sigma'$ and $f_{\Sigma\Sigma}$ becomes unchanged as
seen in the previous section. This fact also indicates the phase space
density is not a covariant expression; if it were covariant, $f_{\Sigma\Sigma'}$
should be unchanged even when $\Sigma\ne\Sigma'$.

Since $f_{\Sigma\Sigma}$ and $f_{\Sigma\Sigma'}$ are the densities
defined on a same spatial volume in $\Sigma$, we can relate them
by\begin{equation}
\frac{1}{E_{\Sigma}}f_{\Sigma\Sigma}\, d\mathbf{p}_{\Sigma'}d\mathbf{x}_{\Sigma}=\frac{1}{E_{\Sigma'}}f_{\Sigma\Sigma'}\, d\mathbf{p}_{\Sigma'}d\mathbf{x}_{\Sigma}\label{eq:lorentz}\,.\end{equation}
 When we apply the above result to the reference measure $\rho$ to
calculate the relative entropy, it has the same meaning as Equation
(34) in Ref \cite{dunkel07}. If the measure $\rho$ is to be invariant
under the transform of $\rho_{\Sigma\Sigma}\rightarrow\rho_{\Sigma\Sigma'}$,
it must be\begin{equation}
\rho(p_{\Sigma})\propto\frac{1}{E_{\Sigma}}\,,\label{eq:Einv}\end{equation}
which is suggested in Ref \cite{dunkel07}. However, as seen from
(\ref{eq:lorentz}), the Lorentz transform in this context is in
the momentum space only and the spatial volume to define the measure
$\rho$ is unchanged.

The present paper proposes that the measure should have the Lorentz
symmetry under the transform both in space and momentum coordinates:
$\rho_{\Sigma\Sigma}\rightarrow\rho_{\Sigma'\Sigma'}$. Then we have
to choose the phase space density defined by (\ref{eq:phi}) instead
of (\ref{eq:phi2}) for the reference measure. As discussed above,
two phase space densities with different reference frames $\Sigma$
and $\Sigma'$ is not directly connected with a equation such as (\ref{eq:lorentz}).
The Lorentz symmetry in this case means the mathematical expression
is unchanged under the transform, and this is satisfied when $N(\bar{x},\bar{p})$
is constant. Therefore we obtain\begin{equation}
\rho(p_{\Sigma})=\textrm{constant}\,,\end{equation}
in the reference frame $\Sigma$ instead of (\ref{eq:Einv}). Following
the relative entropy maximization procedure proposed in Ref \cite{dunkel07}
we obtain the J\"uttner distribution as\begin{equation}
\phi(p_{\Sigma})\propto\exp(-\beta E_{\Sigma})\,.\end{equation}
by maximizing the relative entropy in (\ref{eq:relent}).

\section{Concluding Remarks}

It has been shown in the present paper the maximum entropy state based
on the Lorentz symmetry is the J\"uttner distribution. Recent years
a number of papers have been published claiming the relativistic equilibrium
state is different from the long believed J\"uttner distribution.
Dunkel and coworkers \cite{dunkel07,cubero09} have shed a light to
this controversy by pointing the importance of the reference measure
in the maximum entropy approach.. They have shown that the difference
of the reference measure causes the difference of the equilibrium
distribution as the maximum entropy state. 

Two typical reference measures were suggested in Ref \cite{dunkel07}.
One is constant as a function of $p$ and the other is inversely proportional
to the energy. In Ref \cite{dunkel07} it is conjectured the former
is derived from the invariance of momentum transition, and the latter
comes from the Lorentz symmetry. However, as we have seen in the present
paper, the reference measure with Lorentz symmetry is also found to
be constant when we correctly formulate the covariance of relativistic
phase space density.

The constant reference measure we derived in this paper corresponds
to the constant ``prior probability'' employed by Synge \cite{synge57}.
The information theory was developed long after the days of Synge,
therefore, he did not know the modern concepts such as relative entropy
or reference measure. Nevertheless, his calculation is quite similar
to ours, and the result is the same J\"uttner distribution. (The
author guesses his basic idea historically comes from the probabilistic
interpretation of the entropy by Boltzmann in his late years
\cite{boltzmann78}.) 

Therefore, the argument in the present paper might seem just another
interpretation of Synge's result with information theory if one
believes his derivation. However, considerable number of papers have
been published recently against the J\"uttner distribution and it is
important to clarify the foundation of the maximum entropy process
based on information theory. Moreover, it has become clear in the
present paper what causes the confusion of the reference measure 
(the difference of $\rho_{\Sigma'\Sigma}$ and $\rho_{\Sigma'\Sigma'}$).

The result in the present paper strongly suggests that the relativistic equilibrium
state is the J\"uttner distribution. There are papers in favor of
the J\"uttner distribution in the recent controversy. Debbasch \cite{debbasch08}
critically reviewed the theories proposing alternatives to the J\"uttner
distribution. He examined the relative entropy approach in Ref \cite{dunkel07}
and showed the result would be inconsistent unless the reference measure
is constant. 

Also there is a result of numerical experiment that supports the
J\"uttner distribution \cite{cubero07}.  It was aregued in Ref
\cite{cubero09} that the distribution measured in Ref \cite{cubero07}
is based on what they call ``coordinate-time'', and and the modified
distribution would be obtained if it is defined with
``proper-time''. In this sense, what we examined in the present paper
is the one with ``coordinate-time'', in agreement with the numerical
experiment.

\bigskip{}

It has been known, but has not been well recognized, that the a conserved
quantity (energy-momentum, particle number etc.) distributed over
a finite volume is not a Lorentz invariant quantity because it belongs
to a different time slice of the volume's world tube. Confusions on
this point have caused controversy on the relativistic thermodynamics
(\cite{yuen70,tadas06} and references therein).

To treat this point correctly any spatial density must be expressed
by a flux four vector as a covariant form. The density in the phase
space is no exception. However, the phase space density in the form
of $f(t,\mathbf{x},\mathbf{p})$ is often regarded as a covariant
expression since the value $f$ is unchanged under the Lorentz transform.

As discussed in Section 2, the expression of $f(t,\mathbf{x},\mathbf{p})$
is frame dependent since it is defined on a three dimensional space
volume in a specific reference frame. It seems some of recent papers
do not pay enough attention to this point, and treat the phase space
density in a confusing way. In the present paper we start with the
fully covariant expression of the phase space density \cite{hakim67}
to avoid this confusion. We have seen that the reference measure with
Lorentz symmetry is constant as a function of momentum. Consequently
the maximum entropy state with the Lorentz symmetry is the J\"uttner
distribution. 

It is known that the maximum entropy approach used in Ref \cite{dunkel07}
has the mathematical structure almost parallel to the traditional
ensemble approach \cite{jaynes83}. Therefore, the equilibrium distributions
derived from ensemble approach can be examined with the same basis.
This means the result in the present paper can be applicable to theories
with the traditional approach. 

\bibliographystyle{eplbib}
\bibliography{rel-eq}

\end{document}